\documentclass[12pt,preprint]{aastex}
\def\beq{\begin{equation}}
\def\eeq{\end{equation}}
\def\bey{\begin{eqnarray}}
\def\eey{\end{eqnarray}}

\def\lsim{\mathrel{\raise.3ex\hbox{$<$\kern-.75em\lower1ex\hbox{$\sim$}}}}
\def\gsim{\mathrel{\raise.3ex\hbox{$  $\kern-.75em\lower1ex\hbox{$\sim$}}}}

\begin{document}
\title{Perturbations In A Non-Uniform Dark Energy Fluid: Equations Reveal Effects of Modified Gravity and Dark Matter}
\author{Anaelle Halle\altaffilmark{1,2}, HongSheng Zhao \altaffilmark{1,3}, Baojiu Li\altaffilmark{4}}
\altaffiltext{1}{Scottish University Physics Alliance, University of St Andrews, KY16 9SS, UK, hz4@st-andrews.ac.uk}
\altaffiltext{2}{Ecole Normale Superieure, Ulm, Paris, anaelle.halle@ens.fr}
\altaffiltext{3}{National Astronomical Observatories, Chinese Academy of Sciences, Beijing 100012}
\altaffiltext{4}{DAMTP, Centre for Mathematical Sciences, University of Cambridge}
\begin{abstract}
We propose a unified description of the galactic Dark Matter and various uniform scalar fields for the inflation and cosmological constant by a single field.  The two types of effects could originate from a fluid of both spatially and temporally varying Vacuum Energy if the vacuum has an uneven pressure caused by a photon-like vector field (of perhaps an unstable massive boson).  We propose a most general Lagrangian with a {\bf N}on-{\bf u}niform Cosmological Constant for this vacuum fluid (dubbed as a Nu-Lambda fluid), working within the framework of Einsteinian gravity.  This theory includes a continuous spectrum of plausible dark energy theories and gravity theories, e.g., inflation, quintessence, k-essence, f(R), Generalized Einstein-Aether f(K), MOND, TeVeS, BSTV etc. theories.  It also suggests new models such as a certain f(K+R) model, which suggests intriguing corrections to MOND depending of redshift and density.  
Some specific constructions of the Nu-Lambda fluid (e.g., Zhao's V-$\Lambda$ model) closely resemble the $\Lambda$CDM cosmology on large scale, but fit galaxy rotation curves as good as MOND.  Perturbed Einstein Equations in a simple $f(K_4)$ model are solvable and show effects of a Dark Matter fluid coupled and Dark Energy.  Incorporating the perturbation equations here into standard simulations for cosmological structure growth offers a chance to falsify examples of the Nu-Lambda theories.
\end{abstract}

\date{}


\section{Introducing a framework for vector fields}

General Relativity (GR) is actually a special case and minimal construction of 
a range of theories describing the metric of a plausible universe.  While completely 
adequate on small scales, GR by itself predicts a missing mass and missing energy  
compared to astronomical observations of the metric of the universe on the scale of Kpc to Gpc (e.g., Spergel et al. 2007).
While the missing mass is arguably explained by Dark Matter (DM) particle fields in supersymmetry particle physics, the missing energy almost certainly cannot be explained unless the present universe is immersed in an exotic Dark Energy (DE) field (White 2007).  Since both the effects of DM and DE occur when the gravity is weak,  
one wonders if the underlying fields are tracking the metric field of the gravity (Zhao 2006, 2007, 2008).

Quantum gravity and string theory often predict a non-trivial coupling of some vector field, which violates CPT symmetry satisfied by standard physics (Kostelecky \& Samuel 1989, Kostelecky 2004).  It has been considered by Will \& Nordvedt (1972) that a vector field can be coupled to the space-time metric.  This creates a "preferred frame" in gravitational physics.  A global violation is undesirable, but a local violation is allowed. 
A four time-like vector-field with a non-vanishing time component would select 
a preferred direction at a given space-time coordinate.  It is an aether-like fluid present everywhere, somewhat like a dark energy with some preferred direction.  If  
such a vector coupling to matter is zero or small, then
it can evade current experimental detection (e.g. the CPT violation experiments in Princeton).  There have been an increase of interests about such vectors in recent years, especially works by Kostelecky and coworkers (1989, 2004), Foster \& Jacobson (2006) and co-workers, Lim  (2005) and co-workers, Bekenstein (2004), Sanders (2005), and Zlosnik, Ferreira et al. (2006, 2007) and co-workers.  E.g., Foster \& Jacobson (2006) noted that 
a solar system immersed in a unit time-like vector field (called Einstein-Aether, or AE)
of small enough mass coupling to the metric is apparently consistent
with current measurements of PPN parameters.  
Carroll \& Lim (2004)  noted that such a field can 
have effects on the Hubble expansion.  Inspired by these ideas, several workers
especially Bekenstein (2004), Sanders (2004) and 
Zlosnik et al. (2007)  proposed to extend the application to galaxy scale to use   
it to explain missing matter (Dark Matter).  Many have constrained the theory 
using empirical astronomical data (Famaey \& Binney 2005, Zhao \& Famaey 2006, Zhao 2006), including gravitational lensing (Zhao et al. 2006, Chen \& Zhao 2006, Chen 2008, Angus et al. 2007).  Most recently Zhao (2007) found 
a simple Lagrangian within these frameworks to give rise to the DM-DE effects
of the right amplitude, offering a possible explanation of coincidence of DE scale and DM scale in $\Lambda$CDM cosmology.  The model is dubbed Vector-for-Lambda, or Vacuum-for-Lambda, or ${\rm V}\Lambda$.

Here we propose a {\it very general Langragian} of a non-uniform vector field in the vacuum.  It is effectively a Dark Energy fluid in the GR language.  
This fluid with a {\bf N}on-{\bf u}niform Lambda is dubbed a Nu-Lambda fluid.  
We show its relation with existing theories.  
We isolate a simple case, and give the full field equations.  Most importantly we derive the equations governing perturbation growth in the FRW universe.

To build a covariant theory, one starts with 
the Einstein-Hilbert action used for GR
\begin{equation}
S = \int {d^4 x\sqrt { - g} \left[ {\frac{R}{{16\pi G_N }} } \right]}  + S_M 
\end{equation}
where the light speed is $c=1$, and $G_N$ is the gravitational constant, 
and $g$ is the determinant of the metric $g_{\alpha \beta}$; 
The signature taken here is $(-,+,+,+)$; we do not distinguish Roman $abcd$ and greek $\alpha\beta\gamma\delta$ for four indices.   $R$ is the Ricci scalar, describing the curvature of space-time.  $S_M$ is the matter action that describes the matter distribution.  
Variation of this action with respect to the metric gives the Einstein equations:
\begin{equation}
\frac{{\delta S}}{{\delta g^{\alpha \beta } }} = 0 \Rightarrow G_{\alpha \beta }  =  8\pi GT_{\alpha \beta }^{matter} 
\end{equation}
where $G_{\alpha \beta}=R_{\alpha \beta }  - \frac{1}{2}g_{\alpha \beta } R$ is the Einstein tensor, and $T_{\alpha \beta }^{matter}$: the stress-energy tensor of matter defined by: $\delta S_M  =  - \frac{1}{2}\int {d^4 x\sqrt { - g} T_{\mu \nu } \left( x \right)\delta g^{\mu \nu } \left( x \right)} $. This tensor describes the matter distribution.  In the following we will add new fields and terms to the Einstein-Hilbert action.

\subsection{Vacuum vector field with a dynamic norm}

Denote a vector (Bosonic) field in vacuum by $Z^a$, which has generally a variable or dynamic norm.  
The Lagrangian density of many vector theories can then be casted in the general form\footnote{It is optional to add a new term 
$f_4(\varphi) Z^{a} Z^{b} R_{ab}$, which is related to $f(\varphi,K,J)$ via a full derivative of no effects.}
\begin{eqnarray}\label{glagv}
L &=& (1+f_0(\varphi)) R + f(\varphi,K,J) \\
\varphi^2 & \equiv & Z^aZ_a \\
K & \equiv & K^{ab}_{cd} \nabla_a Z^b \nabla_c Z^d \\
J & \equiv & J^{ab}_{cd} \nabla_a Z^b \nabla_c Z^d,
\end{eqnarray}
where $\varphi^2$ is essentially a scalar field made from the norm of the vector field $Z^a$ without introducing new degree of dynamical freedom, the coefficients $K^{ab}_{cd}$ and $J^{ab}_{cd}$ can be lengthy functions of $Z^a$ and the metric $g_{ab}$ with appropriate combinations of upper and lower index.  For example, the
Lagrangian of the ${\rm V}\Lambda$ model (Zhao 2007) is of the form 
\beq
L=R + f_K(K)+f_J(J)
\eeq
where $K= (Z^a \nabla_a Z^b) (Z^c \nabla_c Z^d)$, 
$J=\delta_b^a \delta_d^c (\nabla_a Z^b)(\nabla_c Z^d)$.   To recover scalar-tensor theories or $f(R)$ theories for Dark Energy, we set coefficients of $K$ and $J$ to zero, so end up with a Lagrangian 
\beq
L= R + f_0(\varphi)R + f(\varphi),
\eeq 
where the vector field $Z^a$ has collapsed into its norm, the scalar $\varphi$.
All these thoeries involve only four degrees of dynamical freedom at maximum.
\footnote{The Lagrangian of the BSTV theory of Sanders (2005) can be casted into a similiar expression but with 5 degrees of freedom in the physical frame, with $L= R + d (\varphi) g^{ab} \nabla_a q \nabla_b q + h(q,\varphi) K - f(q,\varphi) J + 2 V(q,\varphi)$, where $q$ is a new dynamical scalar field.  In the slow-roll approximation, we can neglect the dynamical term by setting $d(\varphi) \sim 0$, eliminate $q$ all together by minimizing the action wrt $q$, hence rewrite $L=R + f(\varphi,K,J)$.  Essentially the function $f$ is a slow-varying scalar in BSTV with its five degrees of dynamical freedom. }
 
As a specific illustration where the coupling coefficients are the simplest, 
the Lagrangian scalar density can be
\begin{eqnarray}\label{lagv}
L \left( Z,g \right) =  R + L_{012} + L_{3} + f_4 L_4 + \sum_{i=5}^{8} a_i L_i + \sum_{i=9}^{\infty} a_i L_i.
\end{eqnarray}
For simplicity consider setting all coefficients $a_i=0$ for $i=9,...\infty$, we have 
\begin{eqnarray}
L_{012} &=& a_0 + a_1 \varphi^2 + a_2 \varphi^4, \qquad \varphi^2 \equiv Z^{\alpha} Z_{\alpha} \\
L_{3} &=& a_3 \varphi^2 R  \\
L_4 &=& Z^{\alpha} Z^{\beta} R_{\alpha \beta}. 
\end{eqnarray}
and
\begin{eqnarray}
L_5 &=& ( \nabla_{\alpha} Z_{\beta} ) ( \nabla^{\alpha} Z^{\beta} )\\ 
L_6 &= & \left(\nabla_{\alpha} Z^{\alpha} \right)^2 \\
L_7 &=& ( \nabla_{\alpha} Z_{\beta} ) ( \nabla^{\beta} Z^{\alpha} )\\ 
L_8 &=& ( Z^{\beta}\nabla_{\beta} Z^{\alpha} ) ( Z^{\gamma}\nabla_{\gamma} Z_{\alpha} )
\end{eqnarray}
where the coefficients $a_i=const$.  

This Lagrangian density can be simplified further; note the $L_4$ term is related to $L_6$ and $L_7$ by a total divergence.  
\begin{eqnarray}\label{fullderiv}
f_4 L_4 &= & \left[ \nabla_a (f_4 W^a) - W^a \nabla_a f_4 \right] - f_4 L_6 - f_4  L_7 \\
W^a &=& \left( Z^a \nabla_b Z^b - Z^b \nabla_b Z^a \right),
\end{eqnarray}
here we can drop the term proportional to $\nabla f_4$, which is zero if $f_4$ is a$\varphi$-independent constant.  
The total divergence term, when integrated over volume, can be dropped in the total action $S$, because the term becomes 
a surface integration over the boundary according to the Stokes theorem.
We can therefore choose not to consider the term $L_4$, absorbing its 
contribution in the $L_6$ and $L_7$ terms.

\subsection{A general Lagrangian of a scalar field plus a unit-vector}

So far $Z^a$ is a vector, not required to be unit-norm. 
It is, however, easier to work with unit vector.  
Now we decompose the vector $Z^a$ into a scalar field $\varphi$ representing its norm 
\beq
Z^{a} = \varphi  \AE^{a}, 
\eeq
pls a unit time-like vector $\AE_{a} = Z_{a}/\varphi$.
Basically 
\beq
Z^{a} Z_{a} \equiv \varphi^{2}, \qquad \AE^a \AE_a=1.
\eeq
Note the sign convention for the unit vector.
 
The time-like unit vector guarantees Lorentz invariance to be broken locally, so that it will always have a non-vanishing time-like component. The additional constraint can be enforced using a non-dynamic Lagrange multiplier $L^*$. 

The co-variant derivative $Z$-terms have the following correspondences to co-variant derivatives of the $\AE$ field and the scalar $\varphi$ field:
\begin{eqnarray}
\nabla_{a}Z_{b}\nabla^{b}Z^{a} = L_5 &=&  \AE^{a}\AE^{b}\nabla_{a}
    \varphi\nabla_{b}\varphi + 2\varphi\nabla_{a}\varphi\AE^{b}\nabla_{b}\AE^{a} + \varphi^{2} K_1 \nonumber\\
 (\nabla_{a}Z^{a})^{2} = L_6 &=& \AE^{a}\AE^{b}\nabla_{a}
    \varphi\nabla_{b}\varphi + 2\varphi\AE^{a}\nabla_{a}\varphi\nabla^{b}\AE_{b}  + \varphi^{2} K_2 \nonumber\\
\nabla_{a}Z_{b}\nabla^{a}Z^{b} = L_7 &=& 
    +g^{ab}\nabla_{a}\varphi\nabla_{b}\varphi + \varphi^{2} K_3 \nonumber\\
(Z^{a}\nabla_{a}Z_{c})(Z^{b}\nabla_{b}Z^{c}) = L_8 &=&  \varphi^{2} \AE^{a}\AE^{b}\nabla_{a}
    \varphi\nabla_{b}\varphi + \varphi^{4}K_4
\end{eqnarray}
where the $K_{i}$s are defined as
\begin{eqnarray}
K_{1} &=& g^{ab}g_{cd}\nabla_{a}\AE^{c}\nabla_{b}\AE^{d}\ =\
\nabla_{a}\AE_{b}\nabla^{a}\AE^{b},\nonumber\\
K_{2} &=&
\delta^{a}_{c}\delta^{b}_{d}\nabla_{a}\AE^{c}\nabla_{b}\AE^{d}\ =\
(\nabla_{a}\AE^{a})^{2},\nonumber\\
K_{3} &=&
\delta^{a}_{d}\delta^{b}_{c}\nabla_{a}\AE^{c}\nabla_{b}\AE^{d}\ =\
\nabla_{a}\AE_{b}\nabla^{b}\AE^{a},\nonumber\\
K_{4} &=& \AE^{a}\AE^{b}g_{cd}\nabla_{a}\AE^{c}\nabla_{b}\AE^{d}\
=\ \AE^{a}\nabla_{a}\AE_{c}\AE^{b}\nabla_{b}\AE^{c}.
\end{eqnarray}
 
Inspired by the above specific case and redefining the coefficients,  
we propose a {\it very general Lagrangian} for a dynamical scalar field $\varphi$ 
coupled with a unit-norm vector field $\AE^a$,  
\begin{eqnarray}\label{glag}
\mathcal{L}(\varphi,\AE) &=& \left[1 + c_{0}(\varphi)\right]R + 2V(\varphi) +
\sum_{i=1}^{4}c_{i}(\varphi)K_{i} +
(\AE^{a}\AE_{a}-1)L^* \nonumber\\
&& + \left[d_{1}(\varphi)g^{ab} + d_{2}(\varphi)\AE^{a}\AE^{b}\right]
\nabla_{a}\varphi\nabla_{b}\varphi\nonumber\\
&& + \left[ d_{3}(\varphi)\AE^{a}\nabla_{a}\AE^{b} +
d_{4}(\varphi)\AE^{b} \nabla_{a}\AE^{a} \right] \nabla_{b}\varphi 
\end{eqnarray}
where $L^*$ is the Lagrange multipler,
$R$ is the Ricci scalar, and c's and d's are now treated as general functions
of the scalar field $\varphi$ and these terms are some kind of dynamical
dark energy; $\varphi$ is a singlet (a real number).
It can be turned non-dynamical if $d_1=d_2=d_3=d_4=0$. 
Models with $d_2=d_3=d_4=0$ are simple quintessence models.

\subsection{Nu-Lambda fluid vs. other theories}

Our general Lagrangian $L(\varphi,\AE)$ for the Nu-Lambda fluid includes the Lagrangian $L(Z,g)$ in eq.~(\ref{lagv}) as a special case.  In fact all coefficients $a_i$ in $L(Z,g)$ could be generalized to functions of $\varphi$.  To see this note that 
$f_4(\varphi) Z^aZ^b R_{ab}$ in eq.~(\ref{fullderiv}) 
would contain a term proportional to 
\beq
W^a \nabla_a f_4 = f'_4 W^a \nabla_a \varphi 
\eeq
and $W^a$ can be casted into $\varphi$ and $\AE$, and the end result 
are terms all included in $L(\varphi,\AE)$.  Likewise, any new terms $a_i L_i$
can be collapsed into $\varphi$ and $\AE$ representation.  For example, 
it was proposed (Ferreira et al. 2007) to set $a_2=a_3=0$ to eliminate the $a_3 R \varphi^2$ coupling term, but include four new terms $a_9 L_9 + a_{10} L_{10} + a_{11}L_{11} +a_{12}L_{12}$ in eq.~(\ref{lagv}), would not lead to new terms in our $L(\varphi,\AE)$.  To see that how these new terms are absorbed, we note that 
\begin{eqnarray}
    Z^{b}Z_{c}\nabla_{a}Z^{c}\nabla_{b}Z^{a} &=& L_9 = 
    \varphi^{2}\AE^{a}\AE^{b}\nabla_{a}\varphi\nabla_{b}\varphi +
    \varphi^{3}\AE^{b}\nabla_{b}\AE^{a}\nabla_{a}\varphi\nonumber\\
   Z^{b}Z_{c}\nabla_{a}Z^{a}\nabla_{b}Z^{c} &=& L_{10} =
    \varphi^{2}\AE^{a}\AE^{b}\nabla_{a}\varphi\nabla_{b}\varphi +
    \varphi^{3}\AE^{a}\nabla_{a}\varphi\nabla^{b}\AE_{b}\nonumber\\
   Z_{c}Z_{d}\nabla_{a}Z^{c}\nabla^{a}Z^{d} &=& L_{11} =
    \varphi^{2}g^{ab}\nabla_{a}\varphi\nabla_{b}\varphi\nonumber\\
    Z^{a}Z^{b}Z_{c}Z_{d}\nabla_{a}Z^{c}\nabla_{b}Z^{d} &=& L_{12} =
    \varphi^{4}\AE^{a}\AE^{b}\nabla_{a}\varphi\nabla_{b}\varphi.\nonumber
\end{eqnarray}
Note the right hand sides are all {\it already included} in our Lagrangian eq.~(\ref{glag}), with a specific assignment of our functions $c_{1, 2, 3, 4}, d_{1, 2, 3, 4}$.
\footnote{Our Langrangian for $Z^a$ is a special case if we choose $c_0 = a_3 \varphi^2$, $c_{1, 2, 3}=a_{5,6,7} \varphi^2,~c_4=a_8\varphi^4,~d_{1}=a_7+a_{11}\varphi^2, d_{2}=a_5+a_6+(a_8+a_9+a_{10}) \varphi^2+a_{12}\varphi^4, d_{3, 4}= 2a_{5,6}\varphi+a_{9,10}\varphi^3$.}  

Bekestein's (2004) TeVeS Lagranian can also be casted (Zlosnik et al. 2006) into that of a pure non-unit norm vector field in the physical metric with a Lagrangian 
\beq
L = R + f_J(J) + f_K(K),
\eeq 
where $f_J(J)=J$ and the functional form of 
$f_K(K)$ is determined the MOND interpolation function.  The variables 
$K$ and $J$ are made of terms 
\beq
K = \sum_{i=5}^{12} k_i(\varphi) L_i, \qquad J = \sum_{i=5}^{12} j_i(\varphi) L_i,
\eeq 
where $k_i$ and $j_i$ are functions of the norm $\varphi$, and $L_i$ are the 8 different combinations of the kinetic terms of the non-unit norm vector $Z^a$, which then reduces 
to our Lagrangian $L(\varphi,\AE)$ using scalar field and the unit vector.  

In fact we claim that our Lagrangian $L(\varphi,\AE)$ is general
enough to include several models in the literature as its special cases:
\begin{enumerate}
\item{GR:} GR corresponds to our model with $b_i=0$, $a_i=0$ except that $a_2$ and $a_0$.  
\item \emph{Scalar-tensor gravity}: when $c_{1, 2, 3, 4}, d_{2, 3, 4} =
    0$ and $d_{1} = 1$ it reduces to the scalar-tensor gravity; if
    furthermore $c_{0} = 0$ it becomes the standard scalar field
    theory for inflation and quintessence etc.
\item \emph{$f(R)$ gravity model}: when $c_{1, 2, 3, 4}, d_{1, 2, 3, 4} =
    0$ we could vary the action with respect to $\varphi$ (\emph{nondynamical}
    now) to have $\frac{\delta\mathcal{L}}{\delta\varphi} = 0
    \Rightarrow c'_{0}(\varphi)R + V'(\varphi) = 0$ where a prime
    means $d/d\varphi$. Solving this equation we get $\varphi(R)$
    so that the action becomes that for the $R+f(R)$ theory.
\item \emph{Einstein-Aether model and $f(\mathcal{K})$ model}: set $c_{0}, d_{1, 2, 3, 4} =0$. 
    When $c_{1, 2, 3, 4}=C_{1,2,3,4}$ are constants and  $V(\varphi) = 0$ one obtains 	Jacobson's 	linear \AE-theory (Jacobson \& Mattingly 2001) with a.   More generally 	assuming $c_{1,2,3,4}=C_{1,2,3,4}\varphi$, we can again vary the action wrt $\varphi	$ and obtain $\sum_{i}C_{i} K_{i} + V'(\varphi) = 0$.  We can solve $\varphi$ as 	function of $\mathcal{K} =C_i K_i$, and write the Lagrangian as $f(\mathcal{K})$ (Zlosnik, Ferreira, Starkman 2007, Zhao 2007).
\item \emph{V-$\Lambda$ model}: This requires a non-dynamical scalar doublet $(\lambda_K, 	\lambda_J)$ (Zhao 2007).  We set $d_{1,2,3,4}=c_{0,1,3}=0$, and $c_2, c_4$ are 	functions of the two independent components $(\lambda_K, \lambda_J)$ of the doublet 	respectively.
\item \emph{TeVeS}: This requires the scalar field to be a
    	doublet $(\varphi, \mu)$, that is, it requires two scalar fields. The $\mu$ field is 	non-dynamical (cf. Bekenstein 2004, Zlosnik, Ferreira, Starkman 2006).  The 	expressions for our functions are lengthy.  
\item \emph{BSTV}: This again requires a scalar doublet $(\varphi,q)$ (Sander 2005).  But 	we set $d_1+d_2=h(q)$ and $d_1=f(q)$, and $d_3=d_4=c_{0,1,2,3,4}=0$.
\end{enumerate}

There are various other special cases, not in the literature.  e.g., 
If we set $d_{1,2,3,4}=0$, 
$c_{1,2,3,4}= \varphi C_{1,2,3,4}$, where $C$'s are constants,
and $c_0(\phi) \ne 0$, 
we can vary wrt $\varphi$ to get an equation of motion, 
$\varphi=\varphi(R,K)$, and eliminate the non-dynamical $\varphi$ and its potential $V(\varphi)$, and then 
cast the Lagrangian as $L = R + F(R,K)$ models.   
This kind of models are simpler than TeVeS, since the unit vector has only   3 degrees of dynamical freedom and there is no scalar freedom.  

\section{The $F(Q)$ models}


The dynamics of our general Lagrangian is very rich.  To be specific, let's 
consider the simpler $F(Q)$ where 
$Q=c_0 M^{-2} R + \mathcal{K}$ models, where     
we redefine $c_i$ as constants for $i=1,2,3,4$, and the $c_0$ term includes a dimensionless (linear) dependance on the Ricci scalar, and redefine variables so that the final total action is
\begin{eqnarray}
S = S_M + {1 \over 16\pi G }\int {d^4 x\sqrt { - g} \left[ R + M^2 F(Q)
 + L^* \left( {A^\alpha  A_\alpha   + 1} \right) \right]}  \\ 
Q \equiv c_0 M^{-2} R + \mathcal{K} 
\end{eqnarray}
where $L^*$ is the Lagrangian multiplier, and
\begin{eqnarray}
\mathcal{K} \equiv M^{-2} K^{\alpha \beta }\mathop{}\nolimits_{\gamma \sigma } \nabla _\alpha  A^\gamma  \nabla _\beta  A^\sigma , \\
K^{\alpha \beta }\mathop{}\nolimits_{\gamma \sigma } = c_1 g^{\alpha \beta } g_{\gamma \sigma }  + c_2 \delta _\gamma ^\alpha  \delta _\sigma ^\beta   + c_3 \delta _\sigma ^\alpha  \delta _\gamma ^\beta   + c_4 A^\alpha  A^\beta  g_{\gamma \sigma }.
\end{eqnarray}

Note that we have replaced the $\AE$ field with $A$ field using the opposite sign convention.  
\beq
A^\alpha  A_\alpha   = -1.
\eeq
We will stick to $A$ field (instead of $\AE$ field) for the rest of the paper.

In the case that $c_0=0$, our action is similar to what was considered by Jacobson and co-workers, except for the non-linear $F$-function.  
Notice that dropping the terms in $c_2$ and $c_4$ and considering $c_3=-c_1$, we find $K^{\alpha \beta }\mathop {}\nolimits_{\gamma \sigma } \nabla _\alpha  A^\gamma  \nabla _\beta  A^\sigma= \frac{c_1}{2} F_{\alpha \sigma} F^{\alpha \sigma}$, where $F_{\alpha \sigma}$ is the antisymmetric Maxwell tensor defined by $F_{\alpha \sigma}=\nabla _\alpha  A_\sigma - \nabla _\sigma  A_\alpha$. This simplification was used by Jacobson and by Bekenstein in TeVeS.  

Models with $c_{1,2,3,4}=0$ and $c_0 \ne 0$ are $F(R)$ theories.   
Models with $c_0=c_4=0$ have been studied by Zlosnik et al. (2007) without giving the full equations.  Here we expand on previous results.  

\subsection{Fields equations for $F(Q)$ models}

Now we proceed to obtain the field equations for models, 
where $Q=c_0 M^{-2} R + \mathcal{K}$.  
What must be borne in mind when carrying out the variations is that the two dynamical degrees of freedom considered are the inverse metric $ g^{\mu \nu }$ and the contravariant  vector field $ A^{\mu} $. The contravariant vector is chosen (and not the covariant one) just because once one has chosen to variate the action w.r.t. $g^{\mu \nu }$, the result of this variation will be simpler seeing the form of $K^{\alpha \beta }\mathop {}\nolimits_{\gamma \sigma }$ because we have :
\begin{equation}
\frac{{\delta A^\mu  }}{{\delta g^{\alpha \beta } }} = 0
\end{equation}
where we used the fact that
\begin{eqnarray}
g_{\mu \rho } g^{\rho \sigma }  = \delta _\mu ^\sigma  \Rightarrow \delta g_{\mu \nu }  =  - g_{\mu \rho } g_{\nu \sigma } \delta g^{\rho \sigma } \\
\rightarrow  \frac{{\delta A_\mu  }}{{\delta g^{\alpha \beta } }} = A^\nu  \frac{{\delta g_{\mu \nu } }}{{\delta g^{\alpha \beta } }} =  - g_{\mu \alpha } A_\beta.
\end{eqnarray}

The vector equation is obtained by varying the action w.r.t. $A^{\mu}$:
\begin{equation}
\frac{{\delta S}}{{\delta A^\alpha  }} = 0 \Rightarrow \nabla _\alpha  (F'J^\alpha  \mathop {}\nolimits_\beta  ) - F'y_\beta   = 2L^* A_\beta  
\end{equation}
which we define $F' = \frac{{dF}}{{d\mathcal{K}}}$ and 
\begin{itemize}
\item $J^\alpha  \mathop {}\nolimits_\sigma$ is a tensor current: $J^\alpha  \mathop {}\nolimits_\sigma   = (K^{\alpha \beta } \mathop {}\nolimits_{\sigma \gamma }  + K^{\beta \alpha } \mathop {}\nolimits_{\gamma \sigma } )\nabla _\beta  A^\gamma   
=2  K^{\alpha \beta } \mathop {}\nolimits_{\sigma \gamma } \nabla _\beta  A^\gamma$, 
due to the symmetry here in K.
\item $y_\beta   = \nabla _\sigma  A^\eta  \nabla _\gamma  A^\xi  \frac{{\delta (K^{\sigma \gamma } \mathop {}\nolimits_{\eta \xi } )}}{{\delta A^\beta  }} =
c_4 M^{-2} A^\alpha  \nabla _\alpha  A_\sigma \nabla _\beta  A^\sigma $.
\end{itemize}

To get the Lagrange multiplier $L^*$, we multiply the equation by $A^{\beta}$ and contract.  Once $L^*$ is known, the equation (which has four components $\beta=0,1,2,3$) yields three constraint equations for the vector.  
Varying the action w.r.t. $L^*$ will give the constraint on the norm:
$A^{\alpha}A_{\alpha} = -1$

For the variation of the action $S = \int {d^4 x\sqrt { - g} L} $ w.r.t. the contravariant metric, one must notice that $\frac{{\delta S}}{{\delta g^{\alpha \beta } }} = \int {d^4 x} \sqrt { - g} \left( {\frac{{\delta L}}{{\delta g^{\alpha \beta } }} - \frac{1}{2}g_{\alpha \beta } L} \right)$ where one uses the fact that: $\delta g = gg^{\mu \nu } \delta g_{\mu \nu }  =  - gg_{\mu \nu } \delta g^{\mu \nu }$, $g$ being the determinant of the contravariant metric.

The symmetry of $K^{\alpha \beta } \mathop {}\nolimits_{\sigma \gamma }$ simplifies the equations:
\begin{eqnarray}
 \frac{{\delta \left( {M^2 F} \right)}}{{\delta g^{\alpha \beta } }} 
  &=& W_{\alpha\beta} 
 + F' [ {Y_{\alpha \beta }  + J^\sigma  \mathop {}\nolimits_\eta  \frac{{\delta \left( {\nabla _\sigma  A^\eta  } \right)}}{{\delta g^{\alpha \beta } }}} ] 
\end{eqnarray}
with 
\beq
W_{\alpha\beta} = ( \Xi R_{\alpha \beta} + g_{\alpha\beta}\nabla\nabla \Xi - \nabla_\alpha
\nabla_\beta \Xi ) , \qquad \Xi \equiv { \partial (M^2 F) \over \partial  R},
\eeq
and
\begin{equation}
Y_{\alpha \beta }  = \nabla _\sigma  A^\eta  \nabla _\gamma  A^\xi  \frac{{\delta (K^{\sigma \gamma } \mathop {}\nolimits_{\eta \xi } )}}{{\delta g^{\alpha \beta} }}.
\end{equation}

The variation of the covariant derivative of the contravariant vector field requires to vary the Christoffel symbol (only): 
\begin{equation}
\frac{{\delta \left( {\nabla _\sigma  A^\eta  } \right)}}{{\delta g^{\alpha \beta } }} = \frac{{\delta \left( {\partial _\sigma  A^\eta   + \Gamma _{\sigma \rho }^\eta  A^\rho  } \right)}}{{\delta g^{\alpha \beta } }} = \frac{{\delta \left( {\Gamma _{\sigma \rho }^\eta  } \right)}}{{\delta g^{\alpha \beta } }}A^\rho  
\end{equation}
And we have $\delta \left( {\Gamma _{\sigma \rho }^\eta  } \right) = \frac{{g^{\eta \tau } }}{2}\left( {\nabla _\sigma  \delta g_{\rho \tau }  + \nabla _\rho  \delta g_{\sigma \tau }  - \nabla _\tau  \delta g_{\sigma \rho } } \right)$,  so one eventually find
\begin{equation}
F'J^\sigma  \mathop {}\nolimits_\eta  \frac{{\delta \left( {\nabla _\sigma  A^\eta  } \right)}}{{\delta g^{\alpha \beta } }} =  - \frac{1}{2} \nabla _\sigma  (\mathcal{F'}(J_{(\alpha } \mathop {}\nolimits_{}^\sigma  A_{\beta )}  - J^\sigma  {}_{(\alpha }A_{\beta )}  - J_{(\alpha \beta )} A^\sigma  ))
\end{equation}
dropping divergence terms which would once more contribute only by boundary terms. The brackets denote symmetrization, ie for instance $J_{\left( {\alpha \beta } \right)}  = \frac{1}{2}\left( {J_{\alpha \beta }  + J_{\beta \alpha } } \right)$.

Putting these together and use 
\begin{equation}
\frac{{\delta A^\mu  A_\mu  }}{{\delta g^{\alpha \beta } }} =  - A_\alpha  A_\beta  
\end{equation}
we find
\begin{equation}
G_{\alpha \beta }  =  8\pi GT_{\alpha \beta }^{matter} + \widehat{T}_{\alpha \beta} - W_{\alpha\beta}
\end{equation}
and $\widehat{T}_{\alpha \beta} ^{}$ is the stress-energy tensor of the vector field
\begin{eqnarray}
\widehat{T}_{\alpha \beta} ^{}  = \frac{1}{2}\nabla _\sigma  (\mathcal{F'}(J_{(\alpha } \mathop {}\nolimits_{}^\sigma  A_{\beta )}  - J^\sigma  {}_{(\alpha }A_{\beta )}  - J_{(\alpha \beta )} A^\sigma  ))  \nonumber \\
- \mathcal{F'}Y_{(\alpha \beta )} + \frac{1}{2}g_{\alpha \beta } M^2 \mathcal{F} + L^* A_\alpha  A_\beta  .
\end{eqnarray}

The above equations are actually true for all $F(R,K)$ models, if $F'$  is  interpreted as $\partial F/\partial \mathcal{K}$.
It is interesting that the effect of the $c_0 R$ term behaves partly 
as a rescaling of the gravitational constant by a factor $(1+ \Xi)$.  
Like in F(R) gravity the value of $\Xi \propto c_0$ must be very small, 
to prevent the term $g_{ab} \nabla^2 \Xi$ in 
the correction source term $W_{ab}$ 
to violate stringent constraints on small scales, e.g., the solar system.
Unless stated otherwise, we will set this source to zero for simplicity.   

\section{Full equations for perturbations of $F(\mathcal{K})$ models in an expanding universe}

\subsection{Metric, Matter, and Einstein Tensor}

With above equations of motion 
we consider a Friedman-Robertson-Walker (FRW hereafter) perturbed metric such that
\begin{equation}
ds^2  =  - (1 + 2 \epsilon \phi )dt^2  + a(t)^2 (1 + 2 \epsilon \psi )(dx^2  + dy^2  + dz^2 )
\end{equation}
A static universe is a special case with all quantities independent of time and $a(t)=1$.  
This fairly general metric is a weakly perturbed form of a homogeneous and spatially isotropic universe 
$ds^2  =  - dt^2  + a(t)^2 (dx^2  + dy^2  + dz^2 )$ by setting $\epsilon=0$;
the unperturbed metric is also spatially flat here, i.e. with no curvature parameter. 
The potentials $\phi$ and $\psi$ are Newtonian gravitational potentials, which are generally non-identical.  This form of perturbed metric 
neglects both tensor mode (gravitational wave) and vector mode
perturbations. 

In the following, the equations are developed in orders of $\epsilon$, but { \it  $\epsilon$ is not kept for a lighter expression}. 

{\it Matter:}
For matter fields, we can take
\begin{equation}
T_{\mu \nu } ^{matter}  = (\rho  + P)u_\mu  u_\nu   + Pg_{\mu \nu } 
\end{equation}
which is the stress-tensor of a perfect fluid without any anisotropic stress, with a density $\rho$, a pressure $P$ and with $u_\mu$ the fluid four-velocity satisfying $g_{\mu \nu} u^\mu u^\nu = -1$. If we consider a non-relativistic fluid, hence neglecting the spatial components of $u_\mu$, then in our metric $u_\mu = (-1- \epsilon \phi, 0,0,0)$. 
We have also
\begin{eqnarray}
 T_{00}^{matter}  = \left( {1 + 2\phi } \right)\rho  \\ 
 T_{0i}^{matter}  = T_{ij}^{matter}  = 0 \\ 
 T_{ii}^{matter}  = a^2 \left( {1 + 2\psi } \right)P 
 \end{eqnarray}

{\it Einstein tensor:}
Up to linear order in $\epsilon$ we find
\begin{itemize}
\item $G_{00} =  3H^2  + 6H\partial _t \psi  - \frac{2}{{a^2 }}\partial_i ^2 \psi$
\item $G_{0i}   =  2(H\partial _i \phi  - \partial _t \partial _i \psi )$
\item $G_{xx}   =  (\mathop a\limits^. \mathop {}\nolimits^2  + 2a\mathop a\limits^{..} )[ - 1 + 2(\phi  - \psi )] + (\partial _y \mathop {}\nolimits^2  + \partial _z \mathop{}\nolimits^2 )(\phi  + \psi ) - 2a^2 \partial _t \mathop {}\nolimits^2 \psi  + 2a\mathop a\limits^. \partial _t (\phi  - 3\psi )$
\item $G_{ij}   =   - \partial _i \partial _j (\phi  + \psi ) \ for \ i \neq j$ 
\end{itemize}

\subsection{Vector field}

We take a homogenous and spatially isotropic universe for the background, so the vector field must, in the background, respect this isotropy for the modified Einstein equations to have solutions, so only the time component can be non zero. The constraint on the norm is $g_{\alpha \beta} A^{\alpha} A^{\beta}=-1$ so in the background, we take $A^{\alpha}=\delta^{\alpha}_0$ and one can then expand it and write: 
\begin{equation}
A_{\mu}=g_{\mu\nu}A^{\nu}=(-1,0,0,0)+ (-\epsilon \phi, \epsilon A_x, \epsilon A_y, \epsilon A_z). 
\end{equation}
The constraint on the total vector $g_{00} A^0A^0 \sim -1$ and $g_{00}=-(1+2\epsilon \phi)$ with the perturbed form of the metric fixes $A^0 \sim 1-\epsilon \phi$ and $A_0 \sim g_{00} A^0 \sim -1- \epsilon \phi$. 

We can also derive $\nabla A$ up to linear order in $\varepsilon$, 
whose non-vanishing components are
\begin{eqnarray}
\nabla_i A_0  =  - H A_i , \qquad 
\nabla_0 A_i - \nabla_i A_0  = \partial_i \phi  + \dot{A_i}, \qquad
\nabla_i A_j  = a^2 \left[ { \dot{\psi}  +  H (1 + 2\psi  - \phi )} \right]\delta_{ij}  + \partial_i A_j.
\end{eqnarray}

Following we carried out the calculations of the Einstein equations up to linear order analytically with this metric.
 { Wherever the expressions become very lengthy, it is helpful to break the expressions into a {\it non-spatial} part without the $A_i$ terms and a {\it spatial} part with $A_i$ terms. }
Note also that any term $F$ and its derivatives $F'$ and $F''$ contain implicitly an unperturbed part and a perturbed part.  Finally we define the shorthands $\alpha \equiv c_1+3c_2+c_3$, and define $\partial_i^2 = \partial_i \partial_i$, where $i=1,2,3$ and $\partial_i$ is the co-moving spatial derivatives and $\dot{X}=\partial_t X$ is the synchronous cosmic time derivative.  

\subsection{Kinetic scalar and $F(\mathcal{K})$}

We decompose the kinetic scalar $\mathcal{K}$ into the 0th, 1st and 2nd order terms.
\begin{eqnarray}
\delta \mathcal{K} &\equiv & \mathcal{K} - {3\alpha H^2 \over M^2} =
{3\alpha H^2 \over M^2 }\left[ -2\phi  + 2H^{-1}\dot{\psi}  + {2\over 3a^2 H} \partial_i A_i  \right]\epsilon +\epsilon^2 \left[{c_4  - c_1 \over a^2 M^2}\left( \partial_i \phi + \dot{A}_i \right)^2 + ... \right], 
\end{eqnarray}
where $...$ includes the 2nd order terms 
$+ {3\alpha \over M^2}\left( { - 4H\psi \partial_t \psi  + 5H^2 \phi ^2  - 4H\phi \partial_t \psi  + \left( {\partial_t \psi } \right)^2 } \right)+ {6c_2 H \over M^2 } \phi \partial _t \phi $ and other lengthy terms shown elsewhere (Halle 2007). 
All 2nd order terms are negligible for the linear perturbation calculations. 
However, the very first term 
${c_4  - c_1 \over a^2 M^2} \left(\partial_i \phi \partial_i \phi \right)$
should be considered for static galaxies, where are in the non-linear regime.

{ The terms $F$, $F'$, $\partial_i F'(\mathcal{K})$ are often involved in the Einstein Equation.  We note they have {\it different orders of magnitude}, given by following 
\begin{eqnarray}
\delta F  &\equiv& F - F\left({3\alpha H^2 \over M^2}\right) = F' \delta \mathcal{K},\\\nonumber
\delta F' &\equiv& F' - F'\left({3\alpha H^2 \over M^2}\right) = F'' \delta \mathcal{K},\\\nonumber
\partial_i F' - 0 &=&  F''(\mathcal{K}) \partial_i \mathcal{K},\\\nonumber
\partial_i \mathcal{K} -0 &=& 
{6 \alpha \over M^2} \left(-H \partial_i \phi + \partial_t \partial_i \psi \right)\epsilon 
\end{eqnarray}
where we have moved the 0th order terms to the LHS.  
In the special case where $\alpha=0$, we find $\delta F',\delta F,\partial_i F',\delta(\mathcal{K})$ are all zero up to the 2nd order.  }

\subsection{The Lagrange multiplier}

The vector equation gives the Lagrange multiplier $L^*  = L^{*A}+L^{*B}$,
where
\begin{eqnarray}
L^{*A} = {3F' \over 1+2\phi} \left[ \left( {c_1  + c_2  + c_3 } \right)(H^2+2H\dot{\psi} ) + c_2 ( -{\mathop a\limits^{..} \over a} + H\dot{\phi} - \partial_t^2 \psi)\right]  
-\dot{F'} (H + 3 \dot{\psi})  - {c_3 \partial_i \over a^2} \left( F'\partial_i \phi  \right) 
\end{eqnarray}
\begin{eqnarray}
L^{*B} =
+ 2\alpha {\dot{a} \over a^3} F'\partial_i A_i  
- 3c_2 {\dot{a} \over a^3 } \partial _i (F'A_i)
- {c_2 \over a^3}\partial_t \left( {a F'\partial_i A_i } \right) 
- {c_3 \over a^2}\partial_i \left( {F'\dot{A}_i } \right) 
\end{eqnarray}
where we could drop a 2nd order term $ - 3c_2 {\dot{a} \over a^3} A_i \partial_i F' - {c_3 \over a^2}\partial_i F' \dot{A}_i $; note that both $\partial_i F'$ and $A_i$ are of 1st order.

\subsection{Perturbed equation of motion of vector field}

For the j-component of the EoM of the vector field, we have 
\begin{eqnarray}
0 &=& {c_4-c_1 \over a} \partial_t \left[ a F' (\partial_j \phi + \dot{A}_j)  \right] + {c_1+c_2+c_3 \over a^2} F' \partial_i\partial_i A_j \\\nonumber
  & & +\alpha \left[ (\partial_j \dot{\psi} - H \partial_j \phi) F' + H \partial_j F' + A_j \partial_t (H F') \right],
\end{eqnarray}
where we have dropped 2nd order terms involving product of $(\partial_i F')$ and another first order quantities $(\Phi, \Psi, A_i)$.   This equation resembles equation B1 and B2 in appendix of Lim (2006).  We consider only the scalar mode here $A_j=\partial_j V$ for $j=1,2,3$. 

\subsection{The spatial off-diagonal terms} 

The EE and vector field stress term with $i \neq j$ satisfies
up to linear order 
\begin{equation}
G_{ij} = -\partial_i \partial_j (\phi + \psi) = 
\widehat{T}_{ij}^{} = - {c_1  + c_3 \over 2a} \partial_t \left[ a F' \left( \partial_i A_j + \partial_j A_i \right) \right].
\end{equation}
We can, as in the static case, identify the Newtonian potentials, $\phi+\psi=0$ in the absence of the anisotropic stress $\widehat{T}_{ij}^{}$, i.e, in the (magnetic) case $c_1+c_3=0$, e.g., GR is such a special case.

\subsection{The $0i$ Cross terms}
 
The 0i component of the stress tensor
\begin{eqnarray}
\widehat{T}_{0j} = {c_4  - c_1 \over a} \partial_t\left(a F' (\partial_j \phi + \dot{A}_j) \right) + \alpha \partial_t \left( F'H \right) A_j + \Delta, 
\end{eqnarray}
where $\Delta =   
+{c_3  - c_1 \over 2a^2}\partial_i \left(F' \left( {\partial_j A_i  - \partial_i A_j } \right) \right)$ is a curl-like term, 
and could be dropped in case of the scalar mode where $A_i = \partial_i V$;  even for vector mode, one can drop the 2nd order term ${c_3-c_1 \over 2a^2} \partial_i F'\left( {\partial _j A_i  - \partial _i A_j } \right) $ safely.   So we have the 0x-component EE
\begin{equation}
+ 2 H \partial_j \phi - 2 \partial_j \dot{\psi}  -  \left( {c_4  - c_1 \over a} \right)  \partial_t (a F' (\partial_j \phi + \dot{A}_j) )
 = -8\pi G \rho u_j \sim 0
\end{equation}
for a non-relativistic matter fluid.

\subsection{The 00th Einstein Equation} 

Replacing $L^*$, we find the ${00}$ component of the vector field stress-energy tensor $\widehat{T}_{00}$ satisfies
\begin{eqnarray}
\widehat{T}_{00} &=&   {c_4 - c_1 \over a^2} \partial_i 
\left[ (\partial_i \phi +\dot{A}_i) F'\right] 
+ \alpha F'H ( 3H  + 6\dot{\psi} + 2a^{-2} \partial_i A_i ) 
- {1 + 2\phi \over 2} M^2 F
\end{eqnarray}
where we could drop a 2nd order term ${c_4-c_1 \over a^2} \partial_i F' \dot{A}_i $.

Thus we can write the 00 Einstein equation
with $T_{00}^{matter}  = \left( {1 + 2\phi } \right)\rho$ as
\begin{equation}
 3(1 - \alpha F')\left( H^2 (1-2\phi)+2H\dot{\psi} \right) 
- {2 \alpha F'H \over a^2} \partial_i A_i  
- {2 \over a^2}\partial_i^2 \psi  - {c_4 - c_1 \over a^2}\partial_i (F' (\partial_i \phi+\dot{A}_i) )  
= \left( 8\pi G \rho -{M^2 F\over 2} \right) 
\end{equation}
where we moved the $F$ term to the RHS, and divided 
the factor $\left( {1 + 2\phi} \right)$.  

\subsection{Spatial Diagonal Equations}

The spatial diagonal terms satisfy, e.g., 
\begin{eqnarray}
\widehat{T}_{xx}= a^2 (1+2\psi) 
\left[ -\alpha {1-2\phi \over a^3} \partial_t (F'a^2\dot{a}) + {M^2 F \over 2} \right]
+ \alpha  a^2\left[ - \dot{F'}\dot{\psi} + F'\left (H \dot{\phi} -6H\dot{\psi} - \partial_t^2 \psi \right) \right] + \Delta^1 + \Delta^2.
\end{eqnarray}
where
\begin{eqnarray}
\Delta^1 \equiv 
- \alpha \left[ H \partial_i (F'A_i)
+ {\partial_t \over 3a}(a F'\partial_i A_i) \right], \qquad
\Delta^2 \equiv  
- {c_1+c_3 \over 3a} \partial_t \left[ a F' (3\partial_x A_x- \partial_i A_i)\right]  
\end{eqnarray}
where the summation over $i=1,2,3$ is implicit, and we could drop a 2nd order term $ - \alpha \mathop H \partial_i F'A_i$.
 
Since for matter $T_{ii}^{matter}  = a^2 \left( { 1 + 2\psi } \right)P$, the modified pressure equation by adding the three spatial diagonal equations becomes
\begin{eqnarray}
\left[ - \left( {1 - 2\alpha F'} \right)H^2  - \left( {2-\alpha F'} \right)\frac{{\mathop a\limits^{..} }}{a} + \alpha \mathop {F'}\limits^. H \right](1-2\phi) + \frac{2}{{3a^2 }}{\partial_i^2 } \left( {\phi  + \psi } \right) 
+ \alpha \left( { 3 H F' + \mathop {F'}\limits^. } \right)\dot{\psi}  
\nonumber \\
+ \left(2 - \alpha F'\right) \left(H (\dot{\phi}-3\dot{\psi})  
- \ddot{\psi} \right)
+ \alpha \left[ H \partial_i (F'A_i)
+ {\partial_t \over 3a}(a F'\partial_i A_i) \right] 
= 8\pi G P + {M^2 F\over 2}
\end{eqnarray}
where we have moved the vector field term to the LHS and divided the factor $(1+2\psi)a^2$ on both sides of Einstein's Equation.

\section{Special Cases}

We have thus obtained the perturbations of the vector field stress-energy tensor and the Einstein equation for an vector field with a Lagrangian involving a general function of the kinetic term $\mathcal{K}$.  
As first check, we recover the linear $F=\mathcal{K}$ model of Lim (2006), and extend it to include a $c_4$ term; this is given in Appendix.   
These perturbation equations are also consistent with Li et al. (2007), which use a very different formulation. 
As a summary of equations and further illustrations, let's consider some more special cases in the context of Dark Matter and Cosmological Constant.
 
Two important quantities of later use are
\beq
\tilde{\lambda} \equiv {c_4 - c_1 \over 2 } {dF(\mathcal{K}) \over d\mathcal{K}}, \qquad
\mu \equiv 1 - \tilde{\lambda} = \sqrt{|\mathcal{K}| \over |\mathcal{K}| + 2}.
\eeq
As it will be evident later on, this choice of $F(\mathcal{K})$ recovers MOND in present-day galaxies.

\subsection{$F(\mathcal{K}_4)$ models with $c_1=c_2=c_3=0$}

The perturbation equations become much simpler if we concentrate on models with a pure $c_4$ term.   By letting $c_1=c_2=c_3=0$, we neglect all contributions of other kinematic terms (one can set $c_4=2$ with no loss of generality).  
Up to the linear order the Lagrange multiplier
\begin{eqnarray}
 L^* = 0.
\end{eqnarray}
We find $\psi=-\phi$ from the spatial off-diagonal EE
\begin{eqnarray}
G_{ij} = -\partial_i \partial_j (\phi + \psi)= \widehat{T}_{ij}^{} =0, ~{\rm for}~ i \neq j.
\end{eqnarray}  
 
Collecting terms in the equation of motion 
and replacing $\psi=-\phi$, and considering only scalar mode $A_i = \partial_i V$ we 
find a time-independent quantity 
\begin{eqnarray}
X_i &\equiv& \left[M a(t)^2\right] \tilde{\lambda} E_i, \qquad
E_i \equiv {1 \over M a(t)} \left[\partial_i \phi + \dot{A}_i \right].
\end{eqnarray}
The quantity $X_i$ behaves as a time-independent co-moving gravitational force of an effective "dark matter" perturbation.  It is the product of a dimensionless "electric" part $E_i$ by a dielectric "suspectibility" part $\tilde{\lambda}$, and a changing scale of gravity $M a(t)^2$; note $E_i$ is similar to the Electric field in a four potential
$(\phi,A_1,A_2,A_3)$ of the radiation field. 

The "suspectibility" ${d \mathcal{F} \over d\mathcal{K}}$, or the
dimensionless $\mathcal{K}$ is an implicit function of 
the modulus of the dimensionless "electric" field $E_i$ through the expression of 
\begin{eqnarray}
\left|E\right| & = &\sqrt{ \mathcal{K} \over  2 },  
\qquad \tilde{\lambda}= {d \mathcal{F} \over d\mathcal{K}},
\end{eqnarray}
where we set $c_4=2$.

The spatial variation of the time-dependent vector field $A_i$ tracks 
the time-independent terms $\partial_i (a\phi)$ and $X_i$ by the EoM constraint
\beq
\dot{A}_i = {1 \over a} \left[ { X_i \over \tilde{\lambda} } - \partial_i (a\phi)\right].
\eeq
Time-wise, as the universe expands, $\mathcal{K}$, ${ X_i \over M a^2}$ all approach 0,  and $\tilde{\lambda}$ approaches a finite value, hence $\dot{A}_i \rightarrow 0$.  

The stress tensors of the non-uniform dark energy fluid are given by
\begin{eqnarray}
\widehat{T}_{0i}    &=& {2 \partial_t X_i \over a} =0\\
\widehat{T}_{0}^{0} &=& \tilde{\Lambda} + {8\pi G \rho_{DM,com} \over a^3} \equiv \tilde{\Lambda} + {8\pi G \rho_{DM,com} \over a^3} \\
\widehat{T}_{x}^{x} &=& - \tilde{\Lambda} \equiv { M^2 F \over 2} ,
\end{eqnarray}
where $\tilde{\Lambda}$ is a time-dependent effective "cosmological constant" term, 
and $\rho_{DM,com}$ is a time-independent effective co-moving "Dark Matter" density. 

The Einstein Equations for $G_{0i}$, $G_0^0$ and ${1 \over 3} G_i^i$ become
\begin{eqnarray}
{2 \over a} \partial_t \partial_i \left(a \phi\right) &=& -8\pi G \rho u_i \sim 0,\\
3H^2 + {2 \over a^2}\partial_i^2 \phi &=& + \tilde{\Lambda} + 
8\pi G \left( {\rho_{DM,com} \over a^3} + \rho \right)  \\
\left[ - H^2  - {2\ddot{a} \over a} \right](1-2\phi) &=& - \tilde{\Lambda} + 8\pi G P.
\end{eqnarray}
where we assumed the source being the non-uniform dark energy stress tensor $\widehat{T}$ plus a non-relativistic baryonic perfect fluid of pressure $P$ and density 
$\rho = (1+\delta) {\bar{\rho}_{com} \over a^3}$, where $\delta$ is a growing overdensity,
and $\bar{\rho}_{com}$ is a constant co-moving mean density of baryons. 

In general the vector field $A_\mu = (-1-\phi,A_1,A_2,A_3)$ in $F(\mathcal{K}_4)$ models simply tracks the space-time metric perturbation $\phi$ and scale factor $a(t)$, which tracks the dominant source, be it radiation or baryonic matter. 
Metric perturbation can be printed in the $A_1,A_2,A_3$ fields even in the absence of baryonic matter.  Note that the effects of the vector field  
is more complex than a change of gravitational constant of a baryon-radiation fluid, where Silk damping can erase perturbations; the vector field is not coupled to photons or baryons directly, hence its perturbations can be passed onto baryons after last scattering.
The $\widehat{T}_{00}$ stress contains a DM-like source term, 
which decays with the redshift as fast as the baryonic density $\bar{\rho} = \bar{\rho}_{com} a^{-3}$, but keeping the effective 
DM-to-baryon contrast time-independent, i.e., 
\beq
{ a^{-3} \rho_{DM,com} \over a^{-3} \bar{\rho}_{com} } 
= ~\mbox{\rm independent of time}, 
\qquad 4 \pi G \rho_{DM,com} \equiv \partial_i X_i. 
\eeq

We can further introduce another parameter for the equation of state parameter defined by 
\begin{equation}
w \equiv { \widehat{T}_{x}^{x} \over \widehat{T}_{0}^{0} }
= \left[1 + { 8\pi G \rho_{DM,com} a^{-3} \over \tilde{\Lambda}     } \right]^{-1}.
\end{equation}
Clearly in case of early universe and CMB, $a^{-3}$ is big, so 
\beq
w =0,
\eeq 
i.e., the equation of the state of the vector field is almost exactly Dark Matter like.  
This is important to understand why the vector field can replace Dark Matter (DM) 
in galaxies.  We will show next that  
the $F(\mathcal{K}_4)$ model is essentially a non-uniform dark energy.  A difference with real DM is that DM density perturbations can grow
in co-moving coordinates while  
the "dark" source term $ {2 \partial_i X_i \over a^3} $ in $F(\mathcal{K}_4)$ corresponds to a static non-uniform density in co-moving coordinates.  
These effective DM and DE are actually coupled through the function $\mathcal{F}$,
i.e., $\tilde{\Lambda}$ is a function of $\rho_{DM,com}/a(t)^2$.  E.g., in the limit of weak field, $\mathcal{K} = 2|E|^2 \rightarrow 0$, $\tilde{\lambda} \rightarrow 1$, 
$\tilde{\Lambda} \sim \Lambda_0 - M^2 \mathcal{K}/2 = \Lambda_0 - (X_1^2+X_2^2+X_3^2) a(t)^{-4}$,
and $\rho_{DM,com} a^{-3} = { k_1 X_1 + k_2 X_2 + k_3 X_3 \over 4 \pi G a(t)^3}$, 
where $(k_1,k_2,k_3)$ is a co-moving wavenumber vector.  Eventually 
when $a(t)\rightarrow \infty$, all effective DM dissipates into effective cosmological constant $\Lambda_0$ for any initial conditions of $X_i$. 

In summary $F(\mathcal{K}_4)$ gravity gives particularly simple equations.
The meaning of these equations has been explored in part in the ${\rm V}\Lambda$ model of Zhao (2007).  The model has the effect of a non-uniform Dark Energy fluid, 
which mimicks galactic Dark Matter, can seed cosmic perturbations, and mimics 
the effect of a coupled Dark Matter and Dark Energy fluid.

\subsection{Homogenous and isotropic universe}

As a second check, we consider the general case of $F' \ne cst$, but in the simple case
of the expanding uniform universe.  
The only non-zero component of the Einstein tensor are
\begin{itemize}
\item $G_{00}=3H^2$
\item $G_{xx}  = G_{yy} = G_{zz} = - (\dot{a}^2  + 2a \ddot{a} )$
\end{itemize}

We have therefore the $00$ modified Einstein equation
\begin{equation}
3\left( {1 - \alpha F'} \right)H^2  + \frac{1}{2}M^2 F = 8\pi G\rho 
\end{equation}
and the modified pressure equation
\begin{equation}
 - \left( {1 - 2\alpha F'} \right)H^2  - 2\left( {1 - \frac{1}{2}\alpha F'} \right){ {\mathop a\limits^{..} } \over a} + \alpha \mathop {F'}\limits^. H - {M^2F \over 2} = 8\pi GP
\end{equation}
These results are identical to that of Zlosnik et al. (2007).  This means simply that the $c_4$ term does {\it not} contribute to the expansion
except for providing a zero point of pressure.  

\subsection{A Possible origin of Cosmological Constant}

{ To see the $c_4$ term can contribute as cosmological constant, let us consider  Hubble expansion in the simple case where we set $\alpha=0$.  For such models $\mathcal{K}=0$.
The equations for expansion become very simple
\begin{eqnarray}
3H^2  + {M^2 F \over 2} &=& 8\pi G\rho \\
-H^2  - 2{ {\mathop a\limits^{..} } \over a} - {M^2F \over 2} &=& 8\pi G P,
\end{eqnarray}
so the equation of state of the vector field is
\beq
w = -1 \qquad ~\mbox{\rm for the Hubble expansion at all redshift}
\eeq

Following Zhao (2007) we set the zero-point $F(\mathcal{K}_{solar})=0$ in 
solar-system like strong gravity regime where $\mathcal{K}_{solar} \sim 10^{16}$, 
since the gravity near the earth's orbit is about $10^8M$, where $M \sim 10^{-10}$m/sec$^2$, so 
\begin{eqnarray}
{M^2F \over 2} = \int_{\mathcal{K}_{solar}}^{\mathcal{K}} 2\tilde{\lambda} d\mathcal{K}.
\end{eqnarray}
Taylor expand in the limit of weak gravity $\mathcal{K} \sim 0$, we have
${M^2 F \over 2} \approx -\Lambda_0 + {M^2 \over c_4} \tilde{\lambda}  \mathcal{K} \approx -\Lambda_0 + {\tilde{\lambda} \over a^2} \partial_i\phi \partial_i \phi$,  which has no first order term, but can have a zero point constant $\Lambda_0$, given by
\begin{eqnarray}
\Lambda_0 = -M^2c_4^{-1} \int_{\mathcal{K}_{solar}}^{0} 2\tilde{\lambda} d\mathcal{K} \sim M^2 \ln \mathcal{K}_{solar},
\end{eqnarray}
where for reasons evidently later we take 
$\mu = 1-\tilde{\lambda} = \sqrt{|\mathcal{K}| \over |\mathcal{K}|+2}$.  
As we will see $\Lambda_0$ 
plays the role of the cosmological constant.  Interestingly 
$\Lambda_0 \sim M^2 \ln 10^{16} \sim 36 M^2 \sim H_0^2$, which is the observed
amplitude of the cosmological constant.  The logarithm factor 
explains why the observed $\Lambda$ is significantly greater than $M^2$.
}

\subsection{Static limit}

As another application, we apply our equations to the regime of quasi-static galaxies.  We set the background expansion factor $a=1$.  In the static limit,  the spatial terms of the vector appear only at second order in all the equations.
$\widehat{T}_{\alpha \beta}^{}$ has no cross-terms (up to linear order), so we find $\psi=-\phi$ from $G_{ij}=0$ equation, 
And the only non-zero component of the Einstein tensor is
\begin{equation}
-G_0^0 = G_{00}  = 2 a^{-2}\partial_i^2 \phi 
\end{equation}

For the vector field we have
\begin{eqnarray}
-\widehat{T}_{0}^{0} & = & a^{-2}\partial_i \left[ 2 \tilde{\lambda} \partial_i \phi \right] - \widehat{T}_{x}^{x}, \\
\widehat{T}_{x}^{x} & = & \frac{1}{2}FM^2 \sim -\Lambda_0,
\end{eqnarray}
where the pressure term $\widehat{T}_{x}^{x}$ 
is generally much smaller than $\widehat{T}_{0}^{0} \sim 8 \pi G \rho$, so the equation of state of the vector field is 
\beq
w \sim 0,
 \qquad ~\mbox{\rm in static galaxies where $|k|^2 \phi  \gg \Lambda_0$.} 
\eeq

From the Einstein 00th equation, and neglect the pressure term, 
we find the modified Poisson equation
\begin{equation}
a^{-2} \partial_i  \left( 2 \mu \partial_i \phi \right) = 8\pi G\rho,
\qquad \mu \equiv 1 - \tilde{\lambda}
\end{equation}
We hence recover eq. (9) of Zlosnik et al. (2007), except that we do not require $c_4=0$. 

{ The above equation resembles the MOND Poisson equation in static limit.
However, MOND also requires for present-day galaxy $\mu \rightarrow \sqrt{y}$ when 
$y \equiv {\partial_i \phi \partial_i \phi \over M^2 a^2} \ll 1$ and 
$\mu \rightarrow 1$ when $y \gg 1$, where we identify $M$ with the MOND critical acceleration $a_0$, i.e, $\sqrt{M} \equiv a_0 \sim 10^{-10}$m/sec$^2$.  With no loss of generality we set $c_4=2, c_1=0$.  
The easiest way to {\it match the MOND function} with $F'$ together is 
to require $\alpha=0$, $y=\mathcal{K}/2$, and 
\beq
\mu=1- {c_4-c_1 \over 2} F'(\mathcal{K})= \sqrt{|\mathcal{K}| \over |\mathcal{K}| + 2}.
\eeq  
The latter corresponds to the standard  $\mu$ function of classical MOND, which fits rotation curves of hundred nearby spiral galaxies extremely well.  }

\section{Possible co-variant dependance of the MONDian behavior on redshift,  environment and history} 

As a final application, we note that 
it is possible to deviate from MOND when we consider galaxy models
with $\alpha =c_1 + 3 c_2 + c_3  \ne 0$ in a non-static universe. 
As before we set $\mu =\sqrt{|\mathcal{K}| \over |\mathcal{K}|+2}$.
However, the kinetic scalar $\mathcal{K}$ is up to 2nd order
\begin{equation}
\mathcal{K} \sim 100 \alpha {H(z)^2 \over H_0^2} + 2 y, 
\qquad y \equiv \frac{1}{M^2a^2} \left( {\partial_i \phi \partial_i \phi } \right),
\end{equation}
where $\mathcal{K}_0 \equiv 3\alpha H_0^2/M^2 \sim 100 \alpha$ for $M \sim H_0/6$, 
Hence we find $\mu = \sqrt{ |y + 50 \alpha H(z)^2/H_0^2| \over 1 + |y + 50\alpha H(z)^2/H_0^2|}$ to depend on redshift.  

Finally, coming back to $F(Q)$ models, the free function  
now depends on $Q=c_0 M^{-2} R+\mathcal{K}$, which depends 
the Ricci scalar, which is crudely speaking the density of the system.
For galaxies in an expanding universe,
\beq
Q = c_0 M^{-2} R + \mathcal{K}  \sim (6 c_0  + 3 \alpha) {H^2 \over M^2} + Q_0, \qquad
Q_0={2 c_0  \partial_i\partial_i \phi  +  2\partial_i\phi \partial_i \phi   \over M^2 a^2}, 
\eeq 
Setting $\alpha= - 2c_0$ we can also 
opt out the $H^2$ term or the redshift-dependancy and make $Q=Q_0$.  
E.g., if the MOND function $\mu = 1 - {c_4 -c_1 \over 2} {dF \over dQ} = \sqrt{Q \over Q +2}$, then MONDian behavior will depend on density.
The zero gravity $\partial_i \phi =0$ region has $\mu \sim Q =Q_0= {2 c_0 \partial_i\partial_i \phi  \over M^2 a^2} \sim {8 c_0 \pi G \rho \over M^2} \sim 100 c_0 \delta_\rho \ll 1$, where $\delta_\rho$ is the over-density over the cosmic mean, and we assume $c_0 \ll 1$. 
So the dark matter effect $\mu^{-1}$ could be bigger in a fluffy galaxy cluster than in a dense galaxy in these models.  
In the solar system $Q$ is big due to high density and strong gravity, hence $\mu=1$, we recover GR like behavior.  
The $F(Q)$ models contain also a correction to the Einstein equation due to  a source proportional to  
$-c_0 W_{\alpha\beta} \sim -c_0 F' R_{\alpha\beta} \sim 0$, where the free function $F' \sim 0-1$.  
This correction can be neglected in the case $c_0 \ll 1$
as in most $F(R)$ gravity models.  

Coming back our general Lagrangian $L(\varphi,A)$ with a dynamical $\varphi$ freedom
if $0=d_1=d_3=d_4$ and $d_2 = 1$.  The term $A^aA^b \nabla_a \varphi \nabla_b \varphi$ creats a quintessence like source term in cosmology but does not contribute to static galaxies.
However, in time-dependant systems, this coupling of $A^a$ and $\varphi$ means that      
the MOND $\mu = \varphi$ in these models has not reached its steady state prediction e.g., $\mu=\sqrt{Q \over Q+2}$.  Instead it must be solved from its own equation of motion 
in an unrelaxed system under merging.  

In short, the co-variant version offers 
new possibilities of tailoring the MOND behavior as function of enviornment and redshift
and history.  These possibilities of covariant dependance of the MOND $\mu$-function
are generally welcome,  
since some of the MOND's worst outliers are with gravitationally lensed galaxy clusters under merging at modest redshift, e.g., the Bullet Clusters at $z=0.3$; clusters have generally lower  density than spiral galaxies, where the empirical formula of MOND applies well.  In this sense, the empirical MOND formula is {\it not} a universal rule,  
and there are a range of possible fundamental rules giving the effects of Dark Matter and Dark Energy.

\section{Conclusion}

We have outlined a framework for studying the dark matter and dark energy effects of a vector field.  We have isolated a few simple cases where the perturbation equations for structure formation are the simplest.  Our equations reduce to the non-linear Hubble equation and the nonlinear Poisson equations in the literature; 
Our simplest model with $c_4 \ne 0 =c_1=c_2=c_3=c_0=d_1=d_2=d_3=d_4$ gives particularly simple Einstein's Equations.  Including other coefficients lead a a range of 
new behaviors in structure formation.  
We itemize our main results as following.

\begin{itemize}   
\item The rotation curves of most spirial galaxies can be 
explained if we adopt the MOND dielectric parameter 
$\mu(\mathcal{K}) = 1- {c_4-c_1 \over 2} F' = \sqrt{|\mathcal{K}| \over |\mathcal{K}+2| }$, where $\mathcal{K} \sim 2 y$, where $\sqrt{y}$ is the gravity measured in units of the acceleration scale $M$.
\item The metric-tracking vector field is described by a four vector $A_a=(-1-\phi,A_1,A_2,A_3)$.  It acts as a dark fluid of certain four-velocity.  This fluid is able to store up perturbations in vacuum in a cold dark matter fashion without being dissipated by photons, hence giving the seed for formation of baryonic structures after the epoch of last scattering (Dodelson \& Ligouri 2006, Zhao 2007).
\item This dark fluid has a non-constant equation of state parameter $w$.
In for the pure $c_4 \mathcal{K}$ case the fluid behaves as a $w=-1$ 
cosmological constant $\Lambda_0$ in Hubble expansion, and $w=0$ dark matter in static galaxies.
\item The small amplitude of vacuum pressure $\Lambda_0 \sim H_0^2$ is explained by the
vector field's pressure in galaxies, if the zero point of the pressure is set at the solar system.  Here $\Lambda_0$ is the maximum pressure difference between very strong and very weak gravity.
\item There are co-variant models $F(R+\mathcal{K})$ models with $\alpha = c_1+3c_2+c_3 \ne 0$, and/or $c_0 \ne 0$ which allows the MOND dielectric function to depend on redshift, density, hence it is no longer a universal rule.  
\end{itemize}

Our perturbation equations can be fairly straightwardly generalised 
by superimposing two  $F(Q_1)$ and $F(Q_2)$ terms together.  E.g.,  in the ${\rm V}\Lambda$ model (Zhao 2007), one replaces $F \rightarrow F(c_4 \mathcal{K}_4) + F_2(J)$
where $F_2 \propto J \propto \mathcal{K}_2$ in the matter-dominated regime.
This $J$-term has effects orthogonal to that of the $\mathcal{K}_4$ term.  It can mimic the effects of dark matter in the Hubble equation, but does not contribute to galaxy rotation curves.

The generality of the equations presented here gives the opportunity of exploring various realistic cases.  
With these it is in principle possible to numerically simulate structure formation and Cosmic Microwave Background to falsify these $F(R+\mathcal{K})$ class of models in the style of Skordis et al. (2006), and Li et al. (2007).  

This work is part of AH's master thesis project in ENS Paris, 
done in collaboration with HSZ at University of St Andrews.  
HSZ acknowledges partial support of PPARC Advanced Fellowship
and National Natural Science Foundation of China (NSFC under grant No. 10428308).

{}

\appendix
\section{Linear Models with $F(\mathcal{K})=\mathcal{K}$}

As a first application of our results, 
we generalize the linear model of Lim (2005) to include a $c_4 \mathcal{K}_4$ term.

We let $F(\mathcal{K})=\mathcal{K}$, hence $F'=1$.  We have 
\begin{eqnarray}
M^2 F(\mathcal{K}) = M^2 \mathcal{K} = {3\alpha H^2 } + {3\alpha H^2 }\left( -2\phi  + 2H^{-1}\partial_t \psi + {2 \over 3a^2 H}\partial_i A_i  \right)\epsilon + \left[{c_4  - c_1 \over a^2}\left( (\partial_i \phi+\dot{A}_i) \right)^2 +  ...\right]O(\epsilon^2)
\end{eqnarray}

The vector equation gives the Lagrange multiplier
\begin{eqnarray}
L^*  (1+2\phi) = 3\left( {c_1  + c_2  + c_3 } \right) (H^2+2H\dot{\psi}) + 3c_2 \left( -{\mathop a\limits^{..} \over a} + H\dot{\phi} - \partial_t^2 \psi \right) - {c_3 \over a^2} \partial_i^2 \phi  \\\nonumber
+ 2(c_1+c_2+c_3) {\dot{a} \over a^3} \partial_i A_i  
- {c_2 +c_3 \over a^2}\partial_t \left( { \partial_i A_i } \right) 
\end{eqnarray}

The ${00}$ component of the stress-energy tensor 
\begin{equation}
\widehat{T}_{00}^{} =   {c_4 - c_1 \over a^2} \partial_i^2 \phi + 3\alpha H^2  
+ 6\alpha H\partial_t \psi  - {1 + 2\phi \over 2} M^2 \mathcal{K}
+ 2\alpha {\dot{a} \over a^3 } \partial_i A_i + 
{c_4-c_1 \over a^2} \partial_t (\partial_i A_i),   
\end{equation}

The spatial diagonal term
\begin{eqnarray}
\widehat{T}_{xx}^{} = -\alpha {1+2\psi-2\phi \over a} \partial_t (a^2\dot{a}) 
+ \frac{1}{2} a^2 (1+2 \psi )  M^2 \mathcal{K}
+ \alpha a^2 \left[ {  - 6H \dot{\psi} + H \dot{\phi} -  \partial _t^2 \psi } \right]
\\\nonumber
- {\alpha \over 3} ( 4H + \partial_t) \partial_i A_i  
- {c_1+c_3 \over 3a} \partial_t \left[ a (3\partial_x A_x- \partial_i A_i)\right]  
\end{eqnarray}

The 0x component of the stress tensor 
\begin{eqnarray}
\widehat{T}_{0x}^{} = \left( {c_4  - c_1 } \right)\left( {\partial _t \partial _i \phi  +  H\partial _i \phi } \right)
+ {c_4  - c_1 \over a} \partial_t\left(a \dot{A}_x\right) + \alpha \partial_t \left( H \right) A_x  
+{c_3  - c_1 \over 2a^2}\partial_i \left( {\partial _x A_i  - \partial _i A_x } \right) 
\end{eqnarray}

The spatial off-diagonal terms 
\begin{equation}
\widehat{T}_{ij}^{} = - {c_1  + c_3 \over 2} (H + \partial_t) \partial_{(i} A_{j)} ,
\end{equation}
where the bracket means symmetric permutation of i and j.

It is reassuring that the above equations agree with those of Lim (2006) if we set $c_4=0$.  This confirms our results up to the linear order in the case that $F'=1$.

\end{document}